\documentclass[atoms,preprints,accept,moreauthors,a4paper]{Definitions/mdpi} 
\usepackage{hhline}
\firstpage{1} 
\pubvolume{xx}
\issuenum{1}
\articlenumber{5}
\pubyear{2025}
\copyrightyear{2025}
\history{Received: ; Accepted: ; Published: date}
\usepackage{graphicx}
\usepackage{epsf}

\DeclareMathAlphabet{\mathsc}{OT1}{cmr}{m}{sc}
\def\testbx{bx}%
\DeclareRobustCommand{\ion}[2]{%
\relax\ifmmode
\ifx\testbx\f@series
{\mathbf{#1\,\mathsc{#2}}}\else
{\mathrm{#1\,\mathsc{#2}}}\fi
\else\textup{#1\,{\mdseries\textsc{#2}}}%
\fi}

\usepackage{caption, booktabs, makecell, threeparttable, siunitx}

 \usepackage{color,colortbl}

\Title{UK APAP R-matrix electron-impact excitation cross-sections for modelling
  laboratory and astrophysical plasma}

\Author{ G. Del Zanna$^{1,2}$ and G. Liang$^{3}$ and J. Mao$^{4}$  and N.R. Badnell$^{5}$}
\AuthorNames{Giulio Del Zanna }

\address{%
  $^{1}$ \quad Department of Applied Mathematics and Theoretical Physics, Centre for Mathematical Sciences, University of Cambridge, Wilberforce Road, Cambridge CB3 0WA, UK \\
 $^{2}$ \quad School of Physics \& Astronomy, University of Leicester, Leicester  LE1 7RH, UK \\
 $^{3}$ \quad  Key Laboratory of Optical Astronomy, National Astronomical Observatories, Chinese Academy of Sciences, Beijing 100101, China
  $^{4}$ \quad Department of Astronomy, Tsinghua University, Beijing 100084, China
   $^{5}$ \quad  Department of Physics, University of Strathclyde, Glasgow G4 0NG, UK
}
  
\corres{Correspondence: gd232@cam.ac.uk}

\abstract{Systematic R-matrix calculations of electron-impact excitation
for ions of astrophysical interest have been
performed since 2007 for many iso-electronic sequences as part of
the UK Atomic Process for Astrophysical Plasma (APAP) network.
Rate coefficients for Maxwellian electron distributions have been provided
and used extensively in the literature and many databases for astrophysics.
Here, we provide averaged collision strengths  to be used to model plasma where electrons are  non-Maxwellian, which often occur in laboratory and astrophysical plasma.
We also provide for many ions new Maxwellian-averaged collision strengths which include important
corrections to the published values. The H- and He-like atomic data 
were recently made available in Mao+(2022). 
Here, we provide data for ions of the
Li-, Be-, B-, C-, N-, O-, Ne-, Na-, and Mg-like sequences. }


\begin{document}
%%%%%%%%%%%%%%%%%%%%%%%%%%%%%%%%%%%%%%%%%%

%%%%%%%%%%%%%%%%%%%%%%%%%%%%%%%%%%%%%%%%%%
% \setcounter{section}{-1} %% Remove this when starting to work on the template.
\section{Introduction}

Most analyses of astrophysical plasma emission rely on accurate atomic data and models.
A large original contribution in terms of theory, codes, and production of a
variety of atomic data for  astrophysical applications started in the 1950's
at University College London by a group of atomic physicists led by
Prof. Mike Seaton FRS. The group became very large and joined forces with a
the group led by Prof. Phil Burke FRS at QUB, to provide over a few decades
a large  set of codes.
Perhaps the most well-known ones are the 
Opacity Project (OP), to calculate opacities,
see \cite{1987JPhB...20.6363S} and the Iron Project (IP), see e.g.
\cite{1993A&A...279..298H}. The Iron Project was mainly tasked to
calculate radiative data and cross-sections for
electron-impact excitation (EIE) of  iron ions using close-coupling calculations 
with the  $R$-matrix method.
Later on, the Iron Project codes were further improved and were used to
calculate atomic data for other elements of astrophysical importance.

Nigel Badnell (NRB)
%\citep[(NRB)][]{obituary}
provided many contributions to these codes 
\citep[cf.][]{1999JPhB...32.5583B}. He also developed since the 1980's
a general-purpose  non-relativistic atomic
structure program {\sc autostructure} \citep{badnell:11} which has been used worldwide to
calculate photoionization cross-sections, radiative and
dielectronic recombination rates, and 
became the standard program to calculate the wavefunctions and radiative data
for the IP work. 

During the early 2000s NRB  started to contribute to a
series of Iron Project papers to calculate EIE cross-sections, see
e.g. \cite{2003A&A...401.1177C,2005A&A...430..331C},
and formed the UK APAP collaboration, which originally included
P.J. Storey, G. Del Zanna and H.E. Mason.
With funding from PPARC/STFC for astrophysical applications,
the UK APAP team  produced   a vast amount of EIE and radiative data.
Aside from specific work on complex iron and nickel ions
\citep[see the review in][]{delzanna_mason:2018}, the bulk of the work
over the years was to calculate data for the main isoelectronic sequences.
An earlier review was given in  \cite{badnell_etal:2016}.

NRB kept at Strathclyde\footnote{http://www.apap-network.org/codes.html}
the main repository of all the OP and IP codes and the main 
 atomic data, which are still  used in virtually all modelling codes for laboratory and
 astrophysical plasma worldwide.
 The EIE and associated radiative data were provided in a compact format for easy
 inclusion in  ADAS\footnote{www.adas.ac.uk}, {\it adf04}. As a set of EIE cross-sections for a single
 ion typically require 5-30 Gb of disk space, Maxwellian-averaged
 rates over a range of temperatures have been provided.

 Over the years, work on assessing the data produced by the UK APAP
 team and benchmarking against laboratory and
 astrophysical spectra  for inclusion in the CHIANTI database\footnote{www.chiantidatabase.org}  
 by GDZ \citep[e.g.][]{2015A&A...582A..56D,chianti_v10}
 identified a few errors for some of the ions.
 Such errors were corrected before inclusion in the CHIANTI database but the
 data on the UK APAP website were not updated.
 We describe below these issues and provide the corrected data.  
 
 However, the main aim of the present work is to provide
 bin-averaged cross-sections for all the data we could rescue.
 They will be fundamentally important for analysing emission from non-thermal
 plasma.
 It is well known that in most laboratory plasma the electron distribution is
 not Maxwellian, even in relatively stable plasma such as those produced by
 Electron Beam Ion Traps.
 It is also well known that  the rate coefficients for any distribution 
 could be recovered in principle, if that could be approximated by a sum of
 Maxwellians.
 On the other hand, it is also well known from kinetic theory that such an
 assumption is often not valid \citep[cf.][]{ljepojevic_burgess:1990}.

 Within solar physics, it is well known that during flares electrons are
 strongly non-thermal, see e.g. the review by  \cite{dudik_etal:2017_review}.
 Recently, evidence that this generally occurs in
 quiescent solar active regions has also been found
 \citep[cf.][]{juraj_etal:2020,delzanna_etal:2022_eis_iris}.
 The possibility  that
non-thermal electrons are present even in planetary nebulae
has been discussed in the literature \citep[see, e.g.][]{nicholls_etal:2012,storey_sochi:2015b},
but clearly there will be astrophysical  plasma where
electrons are non thermal.
In the following sections we briefly  review the methods and the 
atomic data.
We then draw the conclusions.

\section{Methods }

 {\sc autostructure} has been used to get 
 target wavefunctions using radial wavefunctions calculated 
in a scaled Thomas-Fermi-Dirac statistical model potential 
with a set of variational scaling parameters and a 
 configuration-interaction (CI) expansion.
The wavefunctions were used to the scattering calculations but also
to provide separatedly a set of consistent radiative data.
The EIE scattering calculations discussed here were large-scale, using the 
$R$-matrix codes  combined  with the intermediate coupling frame  transformation (ICFT)
method, described by \citet{griffin_etal:98}.
The close-coupling (CC) expansion is typically the same as the CI one of the structure calculation.
The ICFT method first calculates the electron-impact excitation in pure $LS$-coupling and then
transforms into a relativistic coupling scheme via the algebraic transformation of the unphysical
scattering or reactance matrices.

As a consequence, the ICFT method is much faster than the  Breit-Pauli $R$-matrix (BPRM) method
\cite{berrington_etal:95}, the B-spline $R$-matrix (BSR) code \citep[see, e.g. ][]{zatsarinny:2006}, 
or the Dirac atomic $R$-matrix code DARC, originally developed by  P. H. Norrington and I. P. Grant 
\citep[see, e.g.][]{norrington_grant:1981}.
Differences in the results obtained from these codes have been reported in the literature.
However, comparisons of similar-size calculations
for the same ion with the BSR, ICFT, and DARC as e.g. carried out by 
\cite{fernandez-menchero_etal:2017_n_4,delzanna_etal:2019_n_4} found good agreement for the main transitions.
Significant differences  were  found for  weaker transitions
and for those to the higher states. 
Such differences were mainly due to  the structure description
and  correlation effects, rather than due to 
the  different treatment of the relativistic effects in the three codes,
{ as discussed e.g. in  \cite{badnell_etal:2016}}.

%\footnote{http://amdpp.phys.strath.ac.uk/rmatrix/ser/darc/}). We refer readers to \citet{fme16}, \citet{agg17}, and \citet{dza19} for recent comparisons among different $R$-matrix methods and the impact on plasma diagnostics.

The electron collisional excitation rate
coefficient for a Maxwellian electron velocity distribution with an
electron temperature $T_{\rm e}$ in Kelvin is:
\begin{equation}
	C_{ij} ={8.63\times10^{-6} \, \over T_{\rm e}^{1/2}} \,
{\Upsilon_{i,j}(T_{\rm e}) \over g_i} \; \exp \left(- \Delta E_{i,j} / kT_{\rm e} \right)  \quad  [{\rm cm}^3~{\rm s}^{-1}]
%\label{c-ij}
\end{equation}
where   $\Delta E_{i,j} = E_j-E_i$, the energy difference between the lower and
upper states $i$ and $j$, $k$ is Boltzmann's constant, 
$g_i$ is the statistical weight of the initial level, and 
$\Upsilon_{i,j} (T_{\rm e})$ is the dimensionless thermally-averaged collision strength
{ (or effective collision strength)}:
\begin{equation}
\Upsilon_{i,j}(T_{\rm e}) = \int_0^\infty
\Omega_{i,j} \exp\left(-{E_{j} \over kT_{\rm e}}\right) d
\left({E_j \over kT_{\rm e}}\right)  \quad ,
\label{u-ij}
\end{equation}
where $E_j$ is the energy of the scattered electron
relative to the final energy state of the ion, and
$\Omega_{i,j}(E)$ is the  dimensionless  collision strength.  
%In what follows, sometimes we refer to the $\Upsilon$ as a {\it rate coefficient}. 

The collision strengths are normally calculated with a fine grid of
incoming electron energy at lower energies, and a coarser grid otherwise,
for a total of typically several thousands of bins. 
 {\sc autostructure} has also been used to calculate the
 high-energy limits for the collision strengths, following \cite{burgess_etal:97} and \cite{chidichimo_etal:03}.
 These limits are added to the collision strengths to form an OMEGA file.
 The  FORTRAN program {\it adasexj.f} is then used to calculate the { effective collision strengths $\Upsilon$}
 over a range of temperatures.
The  collision strengths are extended to high
 energies by interpolation using  the  high-energy limits in the 
\cite{burgess_tully:92} scaled domain before integration.  
 These high-energy limits, the { effective collision strengths $\Upsilon$} and the radiative data
 have been provided in the compact {\it adf04} format.
 
The OMEGA file for each ion  typically occupies 5-30 Gb of disk space in a binary compressed form, hence  makes it hard to publish.
NRB developed and modified over the years the  program {\it adasexj.f} to
read the large collision strength  files  and integral-average them
over a number of bins, so they could be published.  

We provide, together with the attached data, the last public version 
3.17 (22 Oct 2019) of {\it adasexj.f} which we used.
The program produces a {\it adf04\_om} file if the collision strengths are 
binned or an {\it adf04} for the { effective collision strengths $\Upsilon$}.
After various tests, it was decided when binning to keep the number of energies
to a relatively small value of 101 bins, which however reproduces
relatively  accurately the rates for thermal electrons. 
The same program can be used to calculate the { effective collision strengths $\Upsilon$} 
for a  few  non-thermal distributions and produce them in  {\it adf04} format.
As an input, the program can
take either a full-resolution collision strength file or the bin-averaged. Details are found in the header of the file.

We have run the {\it adasexj.f} program on the full-resolution collision strength files and
cross-checked if we obtained the same { effective collision strengths $\Upsilon$}  as those
we published for a sample of ions. 
Occasionally, the program encounters for extremely weak transitions
some negative values so it sets them to zeroes.
{\it adasexj} has been modified considerably by NRB over the years, and
different versions can produce slightly different { effective collision strengths $\Upsilon$}. 
The main changes over the years are related to how the
collision strengths are extrapolated to zero energy and to the
high-energy limits, hence typically affect the low and high-temperature
{ values}.

The  {\it adf04} file output of {\it adasexj.f}
contains the $A$-values for the main transitions, as well as the
limit points. For the final production of the {\it adf04} files,
we normally run {\sc autostructure} and calulate all the
$A$-values, also for the weak forbidden transitions, up to at least
a third order multipole.
If users are interested to add these $A$-values to
the new  {\it adf04} files, they can use the program {\it adf04mrgr.f}.
However, we advise the users to search the literature to find
more accurate $A$-values, especially for the ground configurations,
as is usually done for inclusion in the CHIANTI database.
In fact, in most cases the $A$-values calculated by
{\sc autostructure} for the forbidden lines are not very
accurate. Sometimes the published $A$-values have been calculated using
experimental energies and in those cases they are generally accurate.

\section{Atomic data}

\subsection{H-like  and He-like ions}

The earlier  $R$-matrix 
calculations by \cite{whiteford_etal:01} which included 
all the transitions among the 49 levels up to  1s~5$l$
were superseded by the larger calculations described by
\cite{mao_etal_hlike:2024} where { CI and CC } expansions of all the 
configurations up to $n = 6$ were included
(36 states for the H-like ions and 71  for the He-like ions).
The \cite{mao_etal_hlike:2024} calculations included 
radiation damping  which is an important effect for H- and He-like ions, 
as discussed e.g. in \cite{gorczyca_badnell:1996,griffin_ballance:2009}.
The { effective collision strengths $\Upsilon$} as well as the bin-averaged collision
strengths have been made available via ZENODO\footnote{https://doi.org/10.5281/zenodo.7226828}.

As a side issue, we note that \cite{fernandez-menchero_etal:2016_h_he}
showed that excitation rates to 
levels higher than $n=5$ can  be estimated more
accurately with extrapolation procedures, rather than with actual calculations.

\subsection{Li-like ions}

\cite{liang_badnell:2011} used the radiation- and Auger-damped 
ICFT $R$-matrix approach to calculate EIE of all Li-like ions from Be$^+$ to Kr$^{33+}$.
The targets included 204 close-coupling (CC) levels, with valence electrons up to $n=5$ 
and core-electron excitations up to $n=4$.

\begin{figure}
\centerline{
  \includegraphics[width=7.5 cm, angle=0]{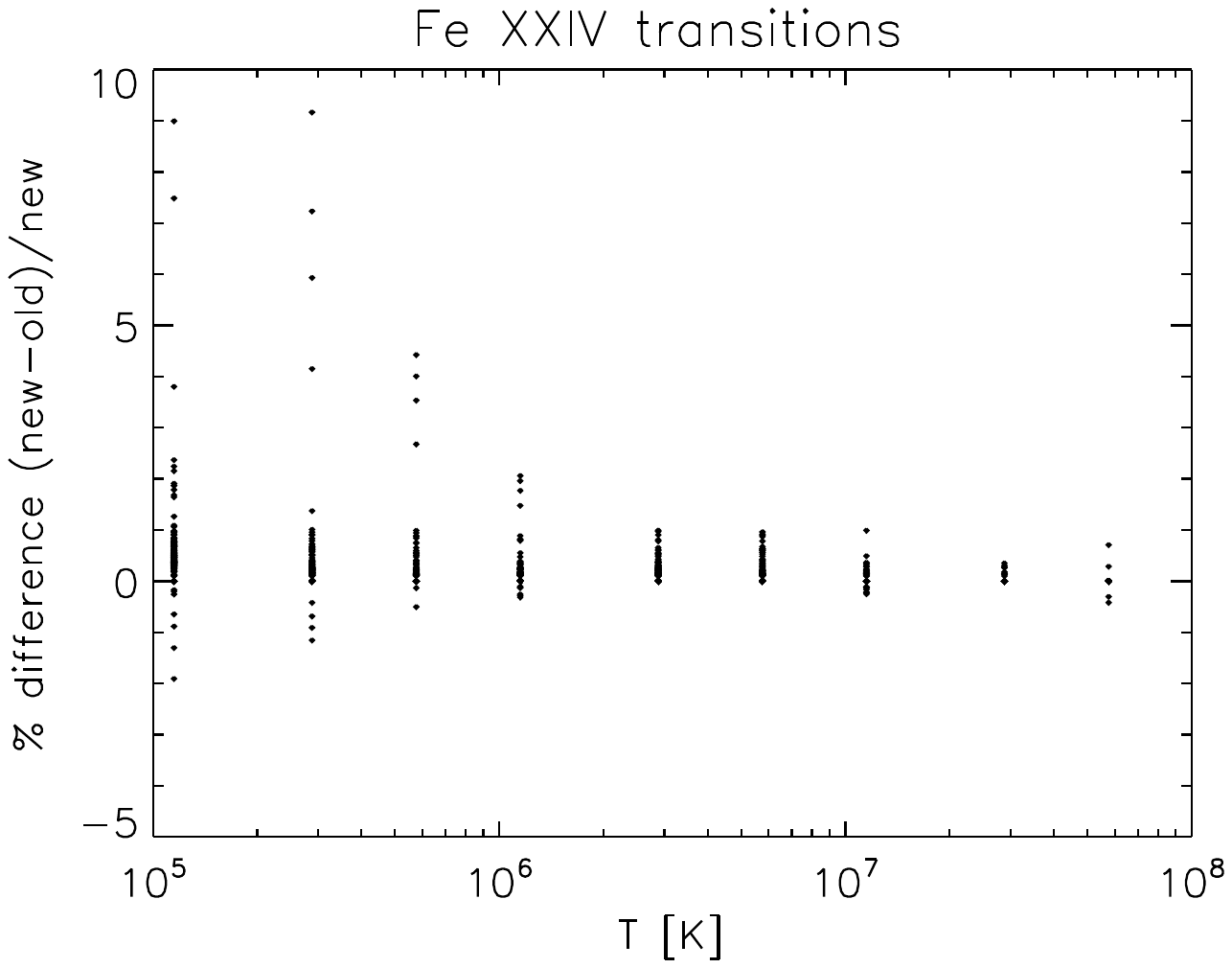}
  \includegraphics[width=7.5 cm, angle=0]{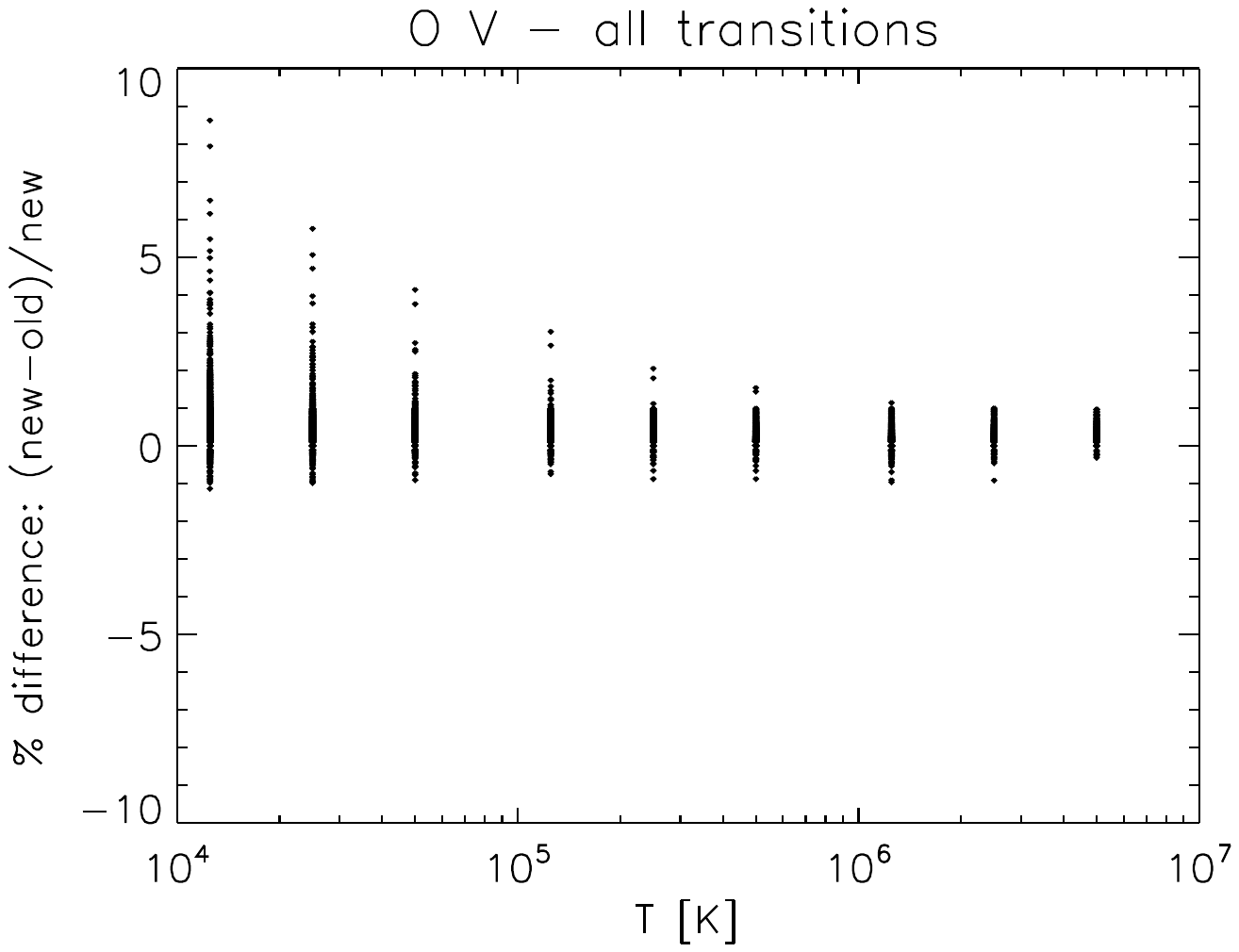}}
\caption{Percentage difference between the present  { effective collision strengths $\Upsilon$}
  (new) and the older ones for Li-like iron (left) and Be-like oxygen (right).}
\label{fig:1}
\end{figure}

During the assessment for CHIANTI v.8 \citep{delzanna_chianti_v8},
small inconsistencies were found for a few ions 
between the highest temperature { effective collision strengths $\Upsilon$}  and  their high-temperature limits.
These were caused by  a mistaken repetition of the last few  collision strengths. 
The data were corrected by GL. 
Some of the corrected ions were included by GDZ in CHIANTI v.8.
Here, we provide the bin-averaged collision strengths
  for the outer shell calculations.  The entire dataset was processed with the corrected OMEGA files.
  We found that the differences between the corrected { effective collision strengths $\Upsilon$}
 and those reprocessed for Fe are within 1--2 percent,  see Figure~\ref{fig:1} (left). Therefore, we provide the 
the older but corrected values, together with the bin-averaged collision strengths for the 33 ions in this sequence.

% Figure~\ref{fig:lilike_fe_24}

\subsection{Be-like ions}

Following the earlier work on  \ion{Mg}{ix} 
\citep{delzanna_etal:08_mg_9} and \ion{Fe}{xxiii} \citep{chidichimo_etal:05},
the ICFT $R$-matrix method was used by
 \cite{fernandez-menchero_etal:2015_be-like}
 to calculate EIE rates for all the ions between  $\mathrm{B}^{+}$
 and $\mathrm{Zn}^{26+}$ in the sequence.
CI and CC expansions of atomic states up to $nl=7{\rm d}$ were included, for  a total
of $238$ fine-structure levels.

The entire dataset was processed and a few tests indicate  close agreement { (within a few percent)
with the published   effective collision strengths $\Upsilon$}, as shown in 
Figure~\ref{fig:1} (right), so we only provide the bin-averaged collision strengths.

\subsection{B-like ions}

The EIE rates for the  boron-like ions from C$^+$ to Kr$^{31+}$
were calculated by \cite{liang_etal:2012} using  the ICFT  $R$-matrix method.
204 close-coupling levels were included in the target,
following the Fe XX  ion model of \cite{badnell_etal:01}.

During the assessment for CHIANTI v.8 \citep{delzanna_chianti_v8}, 
errors in the published data were found.
The 2s 2p$^2$ $^2$S$_{1/2}$, $^2$P$_{1/2}$
levels (No. 8,9) were inverted by mistake (as the experimental
and theoretical energies had different orderings), hence the 
collision strengths  and $A$-values for transitions
 to/from  these levels were incorrect.
 The data for these transitions were recalculated by GL and
 a few  ions were included in CHIANTI by GDZ.

However, tests showed significant discrepancies in the effective 
collision strengths as shown in 
Figure~\ref{fig:bc_comp_ups_all} (left) for B-like iron.
Therefore, the entire dataset of 31 ions 
was reprocessed with the corrected OMEGA files
 and the thermally-averaged collision strengths made available, together with 
 the bin-averaged ones.

\subsection{C-like}

\begin{figure}
\centerline{
\includegraphics[width=7.5 cm, angle=0]{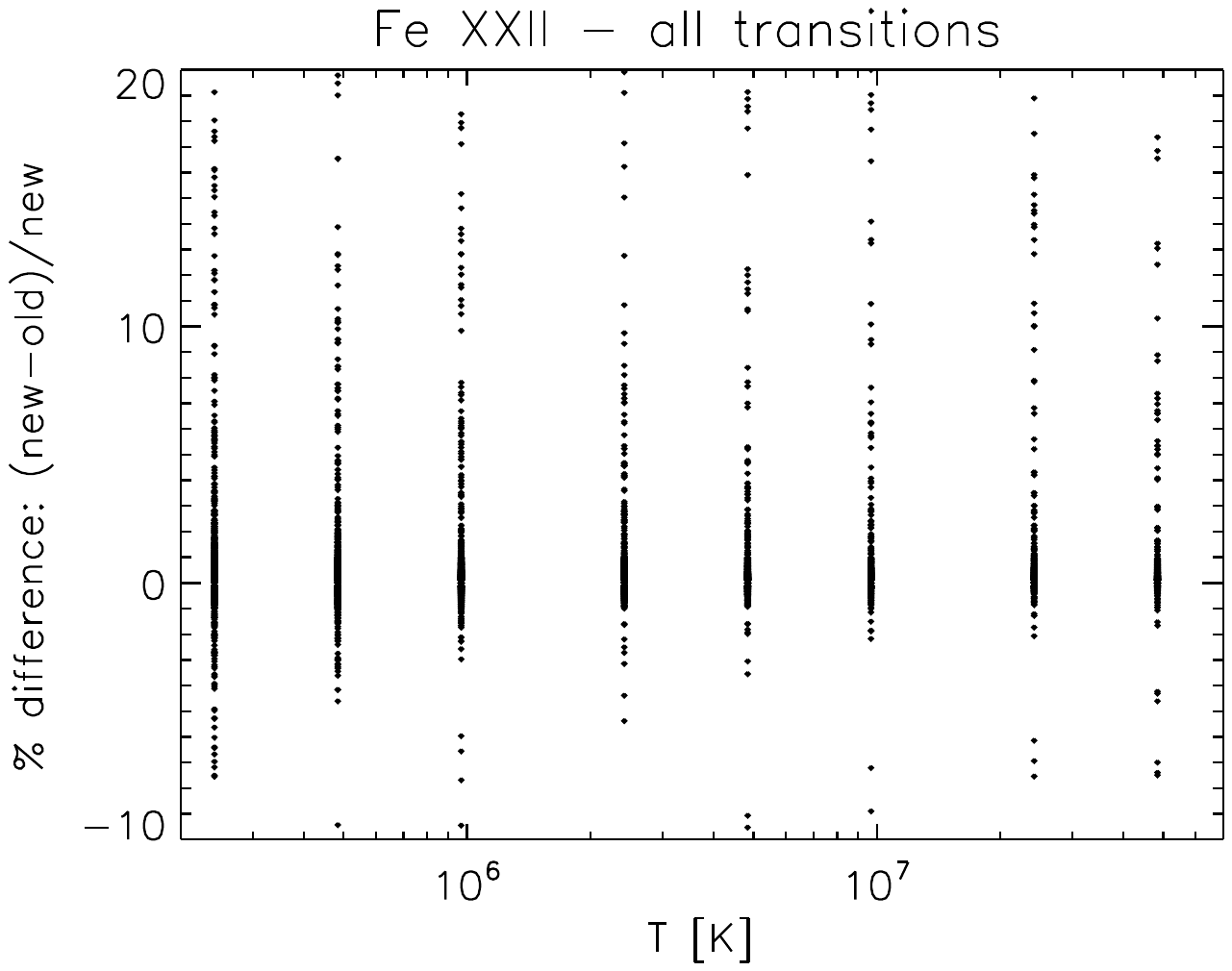}
\includegraphics[width=7.5 cm, angle=0]{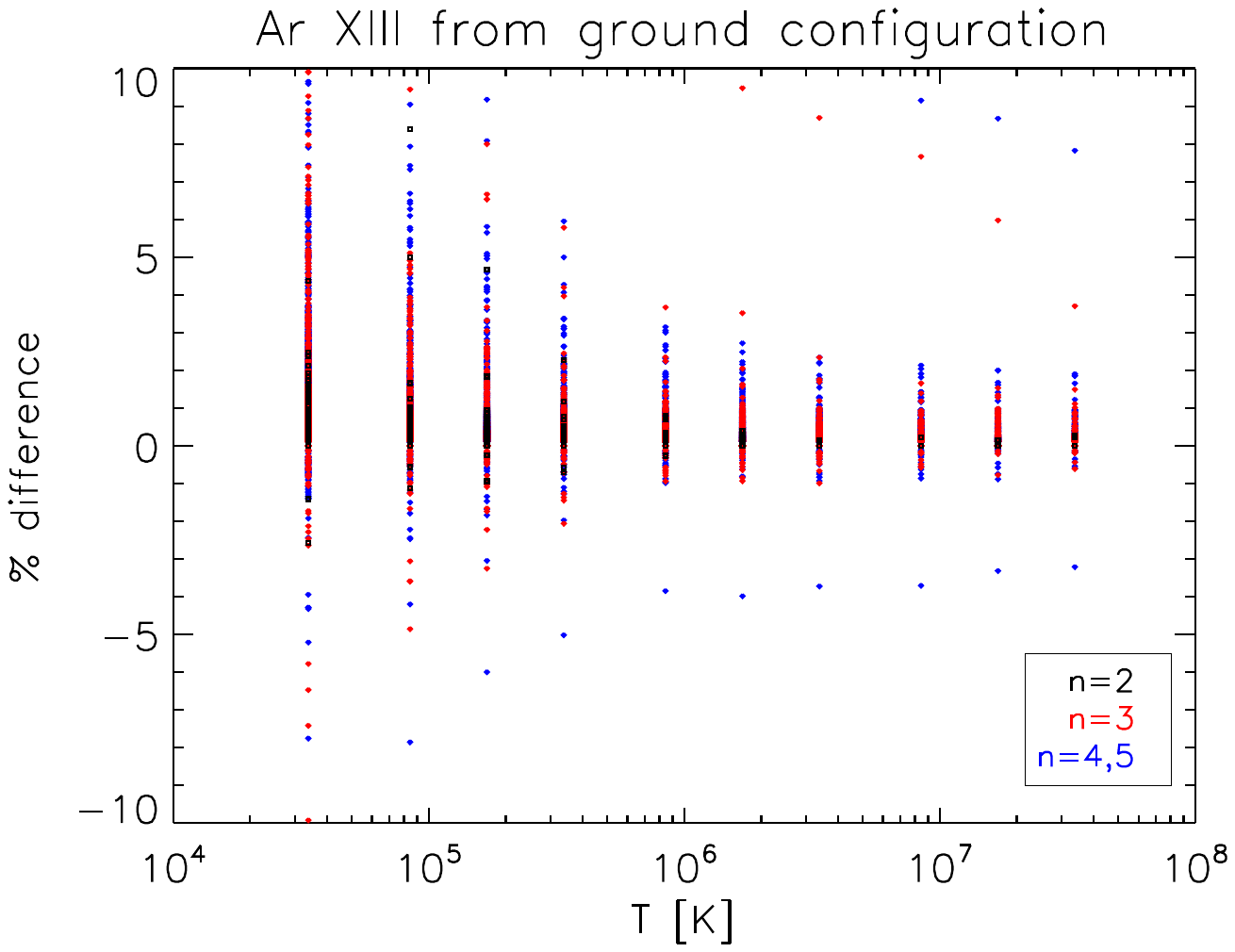}
}
\caption{Left: percentage difference between the present  { effective collision strengths $\Upsilon$}
  and the published ones { (corrected for the level ordering)  for B-like iron.
  Right: the same comparison with the published C-like argon.} }
\label{fig:bc_comp_ups_all}
\end{figure}   
% Figure~\ref{fig:bc_comp_ups_all}

Following the previous study of C-like \ion{Fe}{xxi} by \cite{menchero_etal:2016_fe21},
$R$-matrix ICFT calculations of C-like ions from \ion{N}{ii} to \ion{Kr}{xxxi} (i.e., N$^{+}$ to Kr$^{30+}$)
were reported in \cite{mao_etal:2020}.
A total of 590 fine-structure levels (282 terms) 
were included in the configuration-interaction
target expansion and close-coupling collision expansion for all the ions.
These levels arise from 27 configurations $2l^3 nl^{\prime}$ with $n=2-4$, $l=0-1$, and $l^{\prime}=0-3$
plus the 3 configurations $2s^22p5l$ with $l=0-2$.
A supplementary package can be found at Zenodo\footnote{http://doi.org/10.5281/zenodo.3579183}.
This package includes the inputs for the AUTOSTRUCTURE and $R$-matrix ICFT calculations, as well as the
 { effective collision strengths $\Upsilon$}.

The entire dataset of 29 ions was processed, and differences with the published data was found. We found that the published data were 
processed with an older version, { adasexj\_2.11.f}.
As Figure~\ref{fig:bc_comp_ups_all} (right) shows, the differences are 
only of a few percent for the transitions within the $n=3$ 
states, but become significant for those to the higher levels.
We have therefore recalculated the  { effective collision strengths $\Upsilon$}  for all the ions,
and provide those, together with the bin-averaged collision strengths.

\subsection{N-like ions}

Following the earlier calculations for \ion{Fe}{xx} by \cite{witthoeft_etal:2007}, which included all
the main levels up to $n=4$, \cite{mao_etal:2020_n-like} provided  a larger ICFT calculation involving
CI and CC expansions of 27 configurations containing up to 3 promotions from the
ground configuration $2s^2 2p^3$, up to $n=5$, giving rise to 725 fine-structure states.
All the N-like ions from N$^{+}$ to Zn$^{23+}$ were calculated.
A supplementary package available at Zenodo\footnote{https://doi.org/10.5281/zenodo.4047076}
includes the  { effective collision strengths $\Upsilon$} in {\it adf04} format. 

The entire dataset was processed and a few tests indicate exact agreement
with the published values, so we only provide the bin-averaged collision strengths.

\subsection{O-like ions}

Following an earlier calculation for \ion{Fe}{xix}  by 
\cite{butler_badnell:2008}, where  
the target included 342 close-coupling levels up to $n=4$,
$R$-matrix ICFT calculations for all the O-like ions from Ne$^{2+}$ to Zn$^{22+}$
were reported in \cite{mao_etal:2021_o-like}.
The CI and CC expansions included a total of 630 fine-structure levels
arising from  27 configurations containing up to 3 promotions from the
ground configuration  $2s^2 2p^4$, up to $n=5$.
A supplementary package available at Zenodo\footnote{https://doi.org/10.5281/zenodo.5103521}
includes the input files of the $R$-matrix calculations and the  { effective collision strengths $\Upsilon$}
in {\it adf04} format. 

The entire dataset was processed and a few tests indicate exact agreement
with the published values, so we only provide the bin-averaged collision strengths.

% \subsubsection{F-like ions}

\subsection{Ne-like ions}

\begin{figure*}
\centerline{
\includegraphics[width=7.5cm, angle=0]{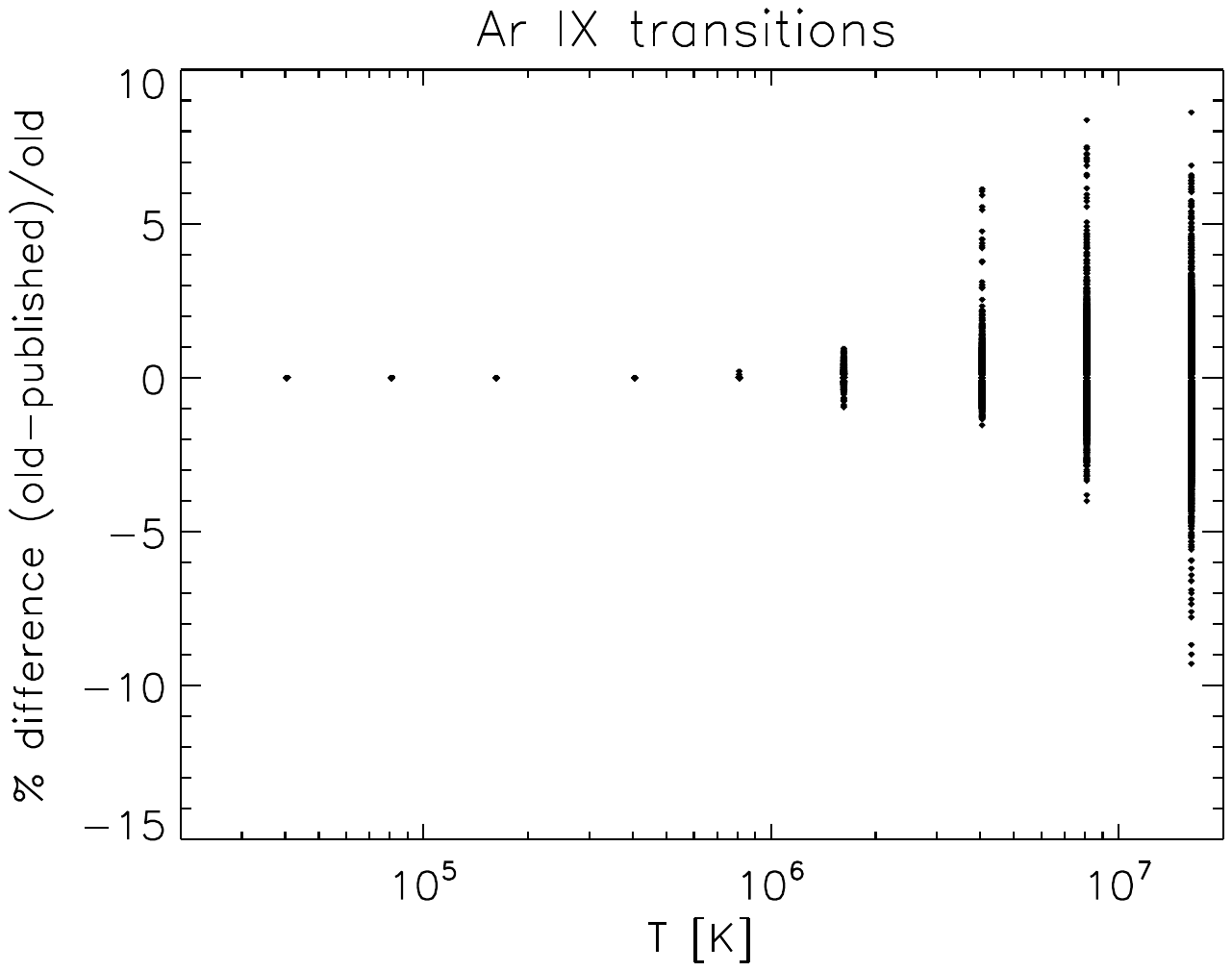}
\includegraphics[width=7.5cm, angle=0]{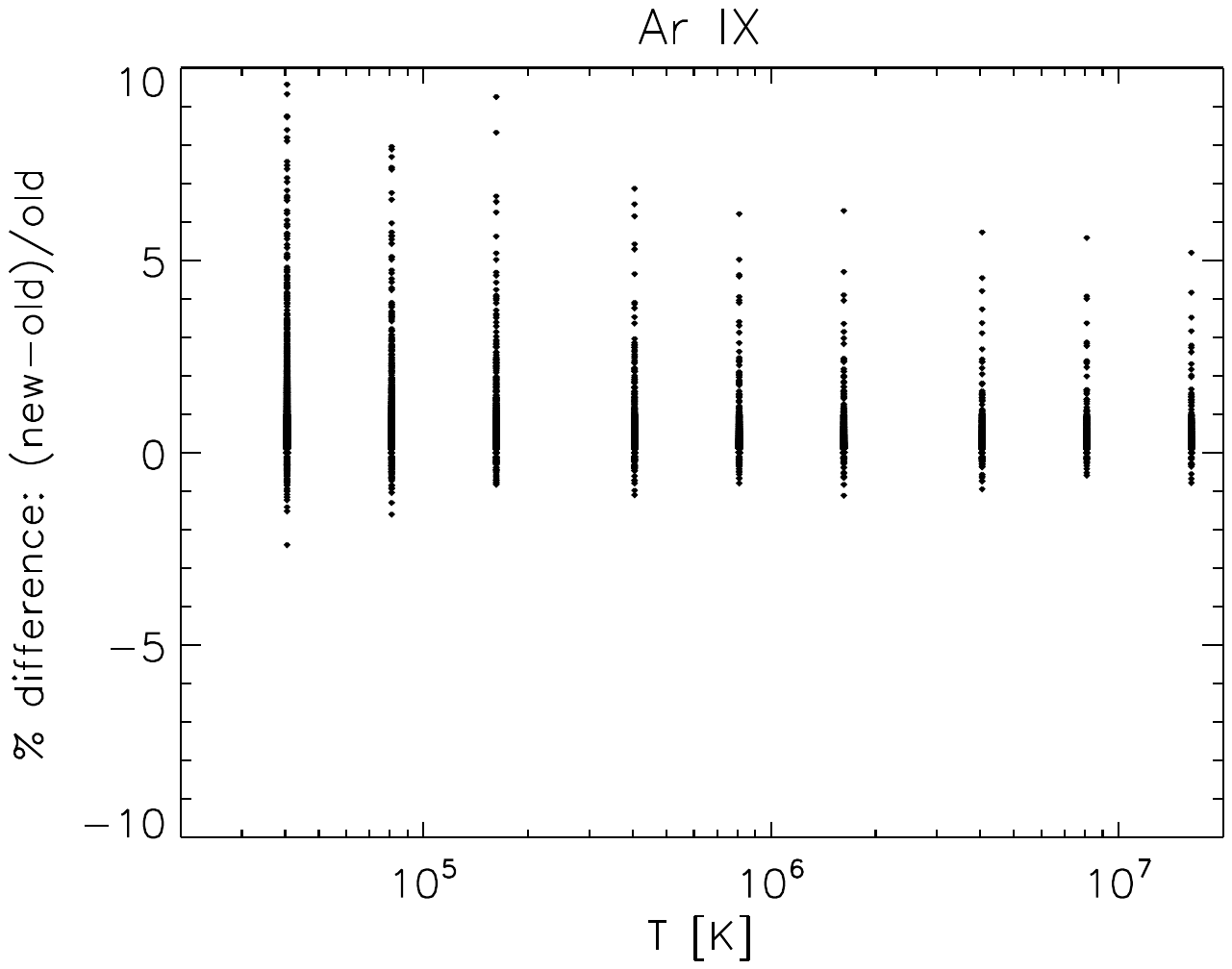}
}
\caption{Left: percentage difference between the corrected { effective collision strengths $\Upsilon$}
  and the published ones for Ne-like argon. Right: 
percentage difference between the present values and the corrected ones. }
\label{fig:nelike_ar_9_comp_ups}
\end{figure*}   
% Figure~\ref{fig:nelike_ar_9_comp_ups}

\cite{liang_badnell:10_ne-like}
calculated with the ICFT $R$-matrix method the EIE { effective collision strengths $\Upsilon$}
for all the ions from Na$^+$ to Kr$^{26+}$. 
The target included  209 levels, up to outer-shell promotions to $n=7$.

As shown by \cite{delzanna:2011_fe_17}, the discrepancies in 
astrophysical observations for the important \ion{Fe}{xvii}  X-ray lines were finally 
resolved whe considering solar flares and either the  
\cite{liang_badnell:10_ne-like} or the Breit-Pauli $R$-matrix calculations
by \cite{loch_etal:06}.

As in the case of the Li-like ions mentioned above, some of the collision strengths 
were affected by the error in the high-temperature values. 
The data  were corrected by GL 
for \ion{Mg}{iii}, \ion{P}{vi}, \ion{Mn}{xvi}, \ion{Si}{v},  \ion{S}{vii},
\ion{Ar}{ix}, and \ion{Ni}{xix} were included in CHIANTI v.8 \citep{delzanna_chianti_v8}.
\ion{Fe}{xvii} and \ion{Kr}{xxvii} were also affected and were fixed. 
As shown in Figure~\ref{fig:nelike_ar_9_comp_ups} (left),
the problem was at temperatures much higher than those where
ions are formed.  

\begin{figure}
\centerline{
\includegraphics[width=7.5cm, angle=0]{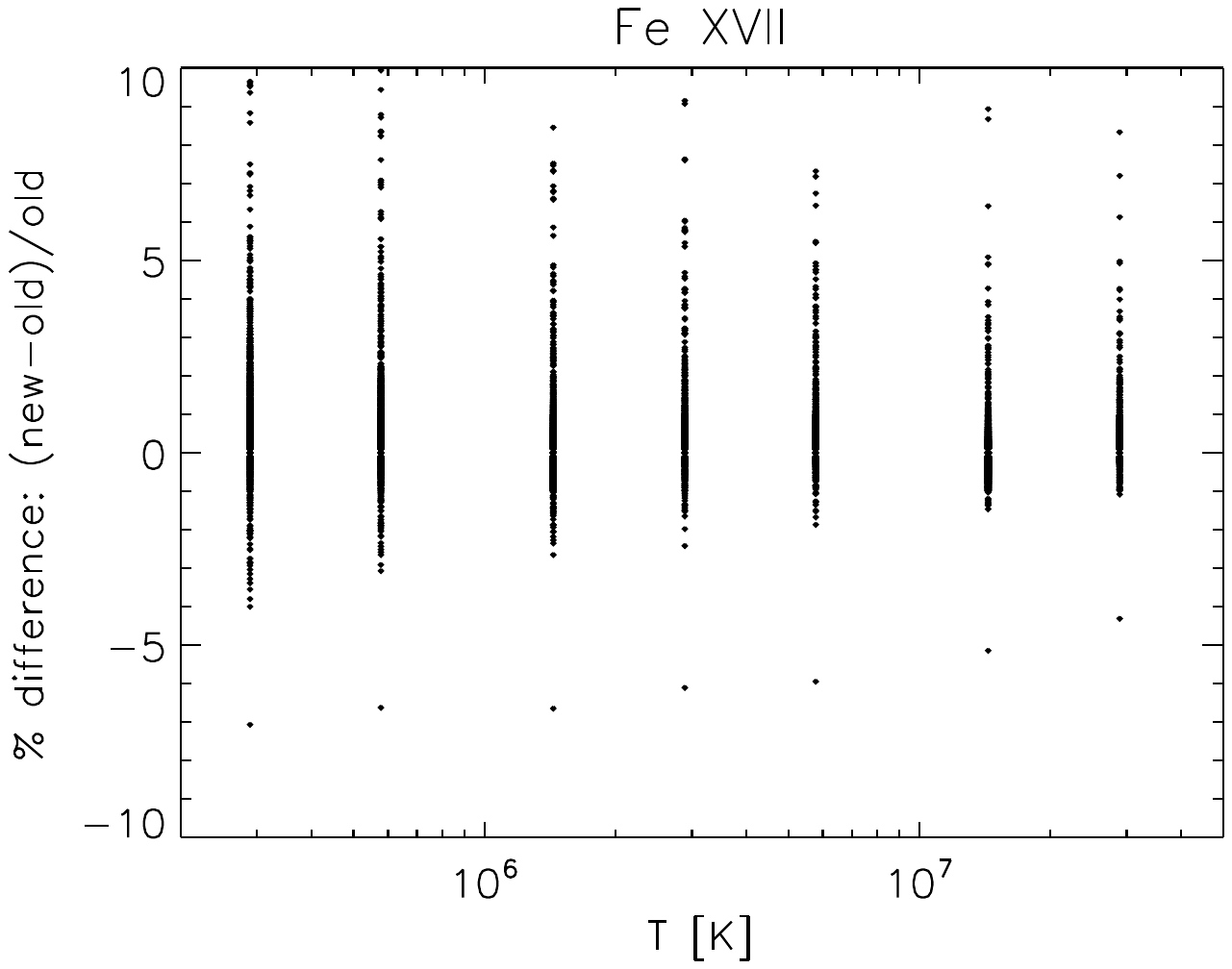}
\includegraphics[width=7.5cm, angle=0]{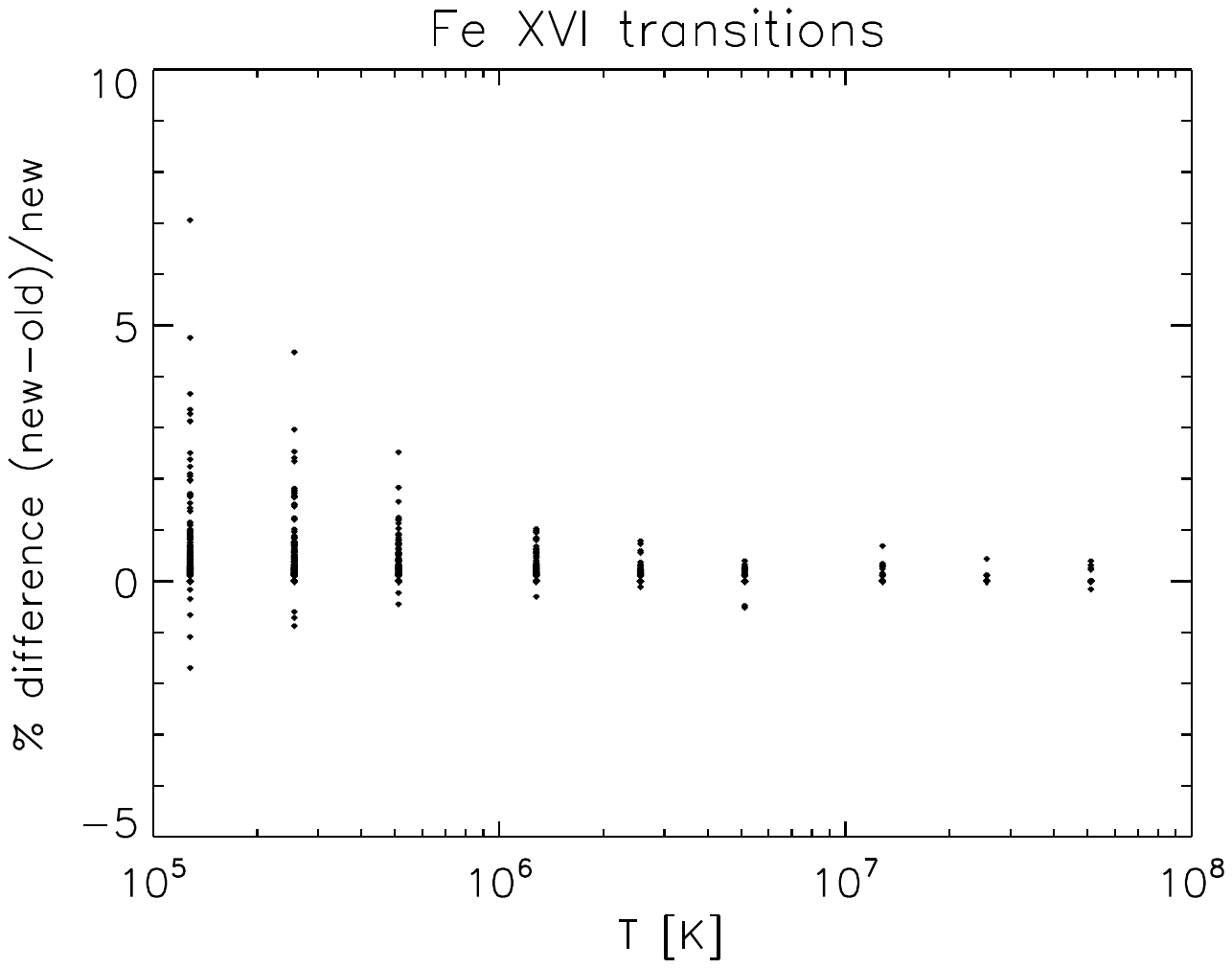}
}
\caption{Percentage difference between the corrected { effective collision strengths $\Upsilon$}
   and the published ones for Ne-like (left) and Na-like (right) iron. }
\label{fig:nelike_fe}
\end{figure}

Figure~\ref{fig:nelike_ar_9_comp_ups} (right) shows 
the difference between the present { effective collision strengths $\Upsilon$} and those that were 
corrected by the high-$T$ mistake.
We also found that the data not affected by the high-$T$ 
problem were processed with the older {\it  adasexj\_2.11.f} version,
with significant differences, see Figure~\ref{fig:nelike_fe} (left).

The entire dataset  was reprocessed and both collision strengths and 
thermally-averaged ones for all the 26 ions are provided.

% Figure~\ref{fig:nelike_fe}

%affected as the  UK APAP website still has the incorrect rates. 

\subsection{Na-like ions}

\cite{liang_etal:09_na-like}  calculated with the ICFT $R$-matrix method the EIE { effective collision strengths $\Upsilon$}
for all the Na-like ions  from  Mg$^+$ to Kr$^{25+}$.
The close-coupling expansion included  configurations up to $n=6$. 
Inner-shell  excitation data with the ICFT $R$-matrix  method 
with both Auger and radiation damping included were 
produced by \cite{liang_etal:09_na-like_inner}.

We are providing the bin-averaged collision strengths
 for the outer shell calculations.  
We found relatively small differences between the present 
and the published { effective collision strengths $\Upsilon$}, as shown in Figure~\ref{fig:nelike_fe} (right).
Therefore, we only provide the collision strengths for the 
25 ions in this sequence.

\subsection{Mg-like ions}

\begin{figure*}
\centerline{
\includegraphics[width=7.5cm, angle=0]{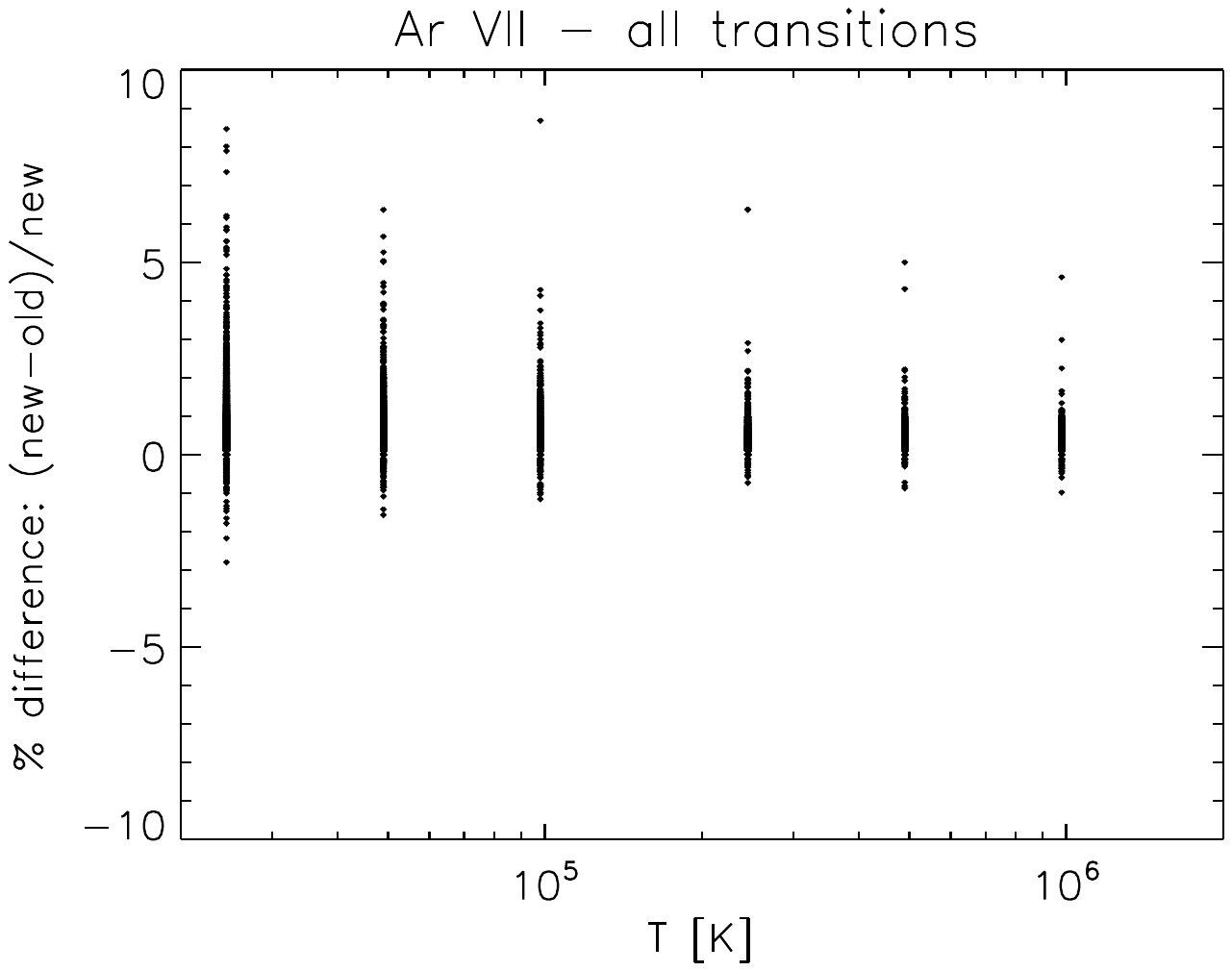}
\includegraphics[width=7.5cm, angle=0]{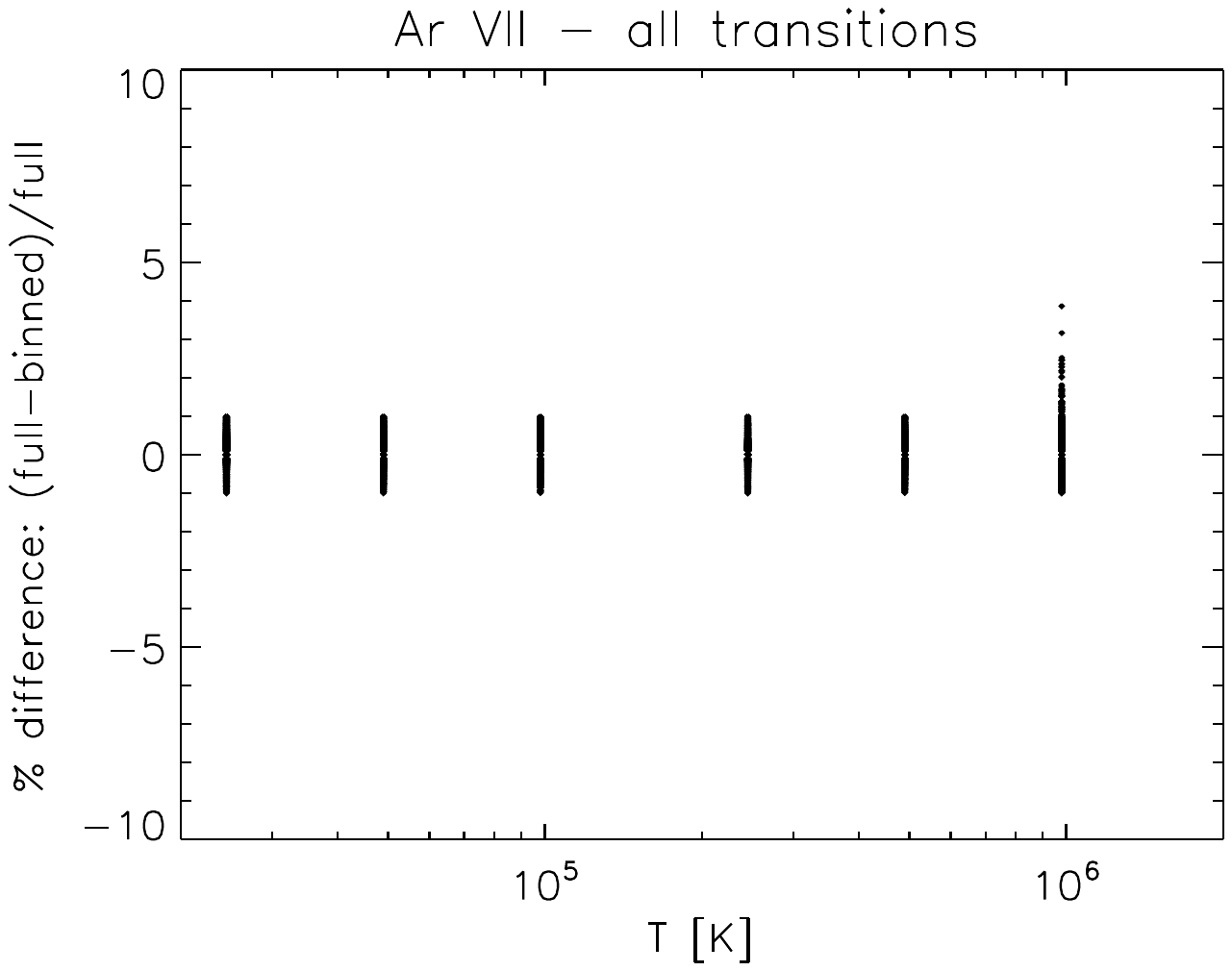}
}
\caption{Left: percentage difference between the present { effective collision strengths $\Upsilon$}
   and the published ones for Mg-like argon. Right: 
percentage difference between the present values and those obtained from the 
bin-averaged cross-sections.}
\label{fig:mglike_ar}
\end{figure*}   
% Figure~\ref{fig:mglike_ar}

 \cite{fernandez-menchero_etal:2014_mg-like} described
ICFT $R$-matrix calculations for all the Mg-like ions from $\mathrm{Al}^{+}$ 
to $\mathrm{Zn}^{18+}$. The target included 
 a total of 283 fine-structure levels in both the 
CI target and CC collision expansions,  from the configurations  
$1\mathrm{s}^2\,2\mathrm{s^2p^6}\,3\{\mathrm{s,p,d}\}\,nl$ with
$n=4,5$, and $l=0 - 4$.

The entire dataset was processed and a few tests indicate 
differences with the published { effective collision strengths $\Upsilon$}, see Figure~\ref{fig:mglike_ar} (left).
We have therefore recalculated all the thermally-averaged collision strengths,
and provide them together with the bin-averaged values 
for all 24 ions in this sequence. 

Finally, in Figure~\ref{fig:mglike_ar} (right) we show that the differences
between the { effective collision strengths $\Upsilon$} obtained from the full OMEGA and those from
the bin-averaged collision strengths are of the order of one/two percent for the temperatures of interest.

\section{Conclusions }

The UK APAP calculations described here are only  part of the work
carried out over the years  under the supervision of Nigel Badnell.
They  have substantially improved on previous work
for all isoelectronic sequences, where in most cases only some
$R$-matrix calculations for a few ions were previous available,
and interpolated data or data of poorer quality was available.

Much of the data described here have already been 
included in various databases and modelling codes to study
laboratory and astrophysical plasma.
The present work fixes several mistakes, provides { effective collision strengths $\Upsilon$}
that have been fixed either during the 
present work or previously. It also provides a
comprehensive set of cross-sections to be used for modelling 
non-thermal plasma.
If possible, we endeavor to process more data and make them available.

The full dataset and associated codes is available on ZENODO 
(DOI: 10.5281/zenodo.14946145) at
https://zenodo.org/records/14946146    

\vskip 0.3 truecm

{\it Ad astra, Nigel}

%%%%%%%%%%%%%%%%%%%%%%%%%%%%%%%%%%%%%%%%%%
\vspace{6pt} 

\authorcontributions{ the original work to run the
  calculations described here was carried out by G. Liang, J. Mao and L. Fern{\'a}ndez-Menchero, under the supervision of Nigel Badnell and  the UK APAP team.  
  The work to rescue the data and process them was carried out
  over a period of six months by GDZ. This work is dedicated to the memory of Nigel 
  Badnell. He was keen to make the present data available. 
  Original draft: GDZ. 
  All authors have agreed to the published version of the manuscript.}

\funding{the present  work by GDZ was unfunded. 
However,  GDZ acknowledges  support from  STFC (UK)  via the consolidated grant
  to the atomic astrophysics group (AAG) at DAMTP,   University of Cambridge (ST/T000481/1).
  The APAP (formerly known as UK Rmax) 
work was funded  by PPARC/STFC (UK) over the past years through the 
University of Strathclyde  grants, with NRB as PI
(1999--2002: PPA/G/S/1997/00783; 
2004--2007: PPA/G/S/2003/0005;2008--2011:PP/E001254/1;
2012--2015: ST/J000892/1; 2018-2021: ST/R000743/1) 
and with various PPARC/STFC grants to partially fund GDZ since 2012.
}

\acknowledgments{We would like to thank the 
University of Strathclyde for the support and in particular Timothy Briggs for maintaining the hardware used by the UK APAP team over the years and helping in finding the disks where the original  work was carried out and the data stored. }

\conflictofinterests{The authors declare no conflicts of interest.}

\reftitle{References}

% Please provide either the correct journal abbreviation (e.g.,according to the “List of Title Word Abbreviations” http://www.issn.org/services/online-services/access-to-the-ltwa/) or the full name of the journal.
%\begin{thebibliography}{999}
%\providecommand{\natexlab}[1]{#1}
%\end{thebibliography}

%% This is file `aastex.cls', 
%\let\jnl@style=\rmfamily 
%\def\ref@jnl#1{{\jnl@style#1}}% 

\newcommand\aj{{AJ}}% 
          % Astronomical Journal 
\newcommand\araa{{ARA\&A}}% 
          % Annual Review of Astron and Astrophys 
\newcommand\apj{{ApJ}}% 
          % Astrophysical Journal 
\newcommand\apjl{{ApJ}}% 
          % Astrophysical Journal, Letters 
\newcommand\apjs{{ApJS}}% 
          % Astrophysical Journal, Supplement 
\newcommand\ao{{Appl.~Opt.}}% 
          % Applied Optics 
\newcommand\apss{{Ap\&SS}}% 
          % Astrophysics and Space Science 
\newcommand\aap{{A\&A}}% 
          % Astronomy and Astrophysics 
\newcommand\aapr{{A\&A~Rev.}}% 
          % Astronomy and Astrophysics Reviews 
\newcommand\aaps{{A\&AS}}% 
          % Astronomy and Astrophysics, Supplement 
\newcommand\azh{{AZh}}% 
          % Astronomicheskii Zhurnal 
\newcommand\baas{{BAAS}}% 
          % Bulletin of the AAS 
\newcommand\jrasc{{JRASC}}% 
          % Journal of the RAS of Canada 
\newcommand\memras{{MmRAS}}% 
          % Memoirs of the RAS 
\newcommand\mnras{{MNRAS}}% 
          % Monthly Notices of the RAS 
\newcommand\pra{{Phys.~Rev.~A}}% 
          % Physical Review A: General Physics 
\newcommand\prb{{Phys.~Rev.~B}}% 
          % Physical Review B: Solid State 
\newcommand\prc{{Phys.~Rev.~C}}% 
          % Physical Review C 
\newcommand\prd{{Phys.~Rev.~D}}% 
          % Physical Review D 
\newcommand\pre{{Phys.~Rev.~E}}% 
          % Physical Review E 
\newcommand\prl{{Phys.~Rev.~Lett.}}% 
          % Physical Review Letters 
\newcommand\pasp{{PASP}}% 
          % Publications of the ASP 
\newcommand\pasj{{PASJ}}% 
          % Publications of the ASJ 
\newcommand\qjras{{QJRAS}}% 
          % Quarterly Journal of the RAS 
\newcommand\skytel{{S\&T}}% 
          % Sky and Telescope 
\newcommand\solphys{{Sol.~Phys.}}% 
          % Solar Physics 
\newcommand\sovast{{Soviet~Ast.}}% 
          % Soviet Astronomy 
\newcommand\ssr{{Space~Sci.~Rev.}}% 
          % Space Science Reviews 
\newcommand\zap{{ZAp}}% 
          % Zeitschrift fuer Astrophysik 
\newcommand\nat{{Nature}}% 
          % Nature 
\newcommand\iaucirc{{IAU~Circ.}}% 
          % IAU Cirulars 
\newcommand\aplett{{Astrophys.~Lett.}}% 
          % Astrophysics Letters 
\newcommand\apspr{{Astrophys.~Space~Phys.~Res.}}% 
          % Astrophysics Space Physics Research 
\newcommand\bain{{Bull.~Astron.~Inst.~Netherlands}}% 
          % Bulletin Astronomical Institute of the Netherlands 
\newcommand\fcp{{Fund.~Cosmic~Phys.}}% 
          % Fundamental Cosmic Physics 
\newcommand\gca{{Geochim.~Cosmochim.~Acta}}% 
          % Geochimica Cosmochimica Acta 
\newcommand\grl{{Geophys.~Res.~Lett.}}% 
          % Geophysics Research Letters 
\newcommand\jcp{{J.~Chem.~Phys.}}% 
          % Journal of Chemical Physics 
\newcommand\jgr{{J.~Geophys.~Res.}}% 
          % Journal of Geophysics Research 
\newcommand\jqsrt{{J.~Quant.~Spec.~Radiat.~Transf.}}% 
          % Journal of Quantitiative Spectroscopy and Radiative Trasfer 
\newcommand\memsai{{Mem.~Soc.~Astron.~Italiana}}% 
          % Mem. Societa Astronomica Italiana 
\newcommand\nphysa{{Nucl.~Phys.~A}}% 
          % Nuclear Physics A 
\newcommand\physrep{{Phys.~Rep.}}% 
          % Physics Reports 
\newcommand\physscr{{Phys.~Scr}}% 
          % Physica Scripta 
\newcommand\planss{{Planet.~Space~Sci.}}% 
          % Planetary Space Science 
\newcommand\procspie{{Proc.~SPIE}}% 
          % Proceedings of the SPIE 
%\let\astap=\aap 
%\let\apjlett=\apjl 
%\let\apjsupp=\apjs 
%\let\applopt=\ao 
%\newcommand\phn{\phantom{0}}% 
%\newcommand\phd{\phantom{.}}% 
%\newcommand\phs{\phantom{$-$}}% 
%\newcommand\phm[1]{\phantom{#1}}% 
%\let\la=\lesssim            % For Springer A&A compliance... 
%\let\ga=\gtrsim 

%% End of file `aastex.cls'. 

\externalbibliography{yes}
\bibliography{paper}

\begin{thebibliography}{-------}
\providecommand{\natexlab}[1]{#1}

\bibitem[{Seaton}(1987)]{1987JPhB...20.6363S}
{Seaton}, M.J.
\newblock {Atomic data for opacity calculations. I. General description}.
\newblock {\em Journal of Physics B Atomic Molecular Physics} {\bf 1987}, {\em
  20},~6363--6378.
\newblock
  doi:{\changeurlcolor{black}\href{https://doi.org/10.1088/0022-3700/20/23/026}{\detokenize{10.1088/0022-3700/20/23/026}}}.

\bibitem[{Hummer} \em{et~al.}(1993){Hummer}, {Berrington}, {Eissner},
  {Pradhan}, {Saraph}, and {Tully}]{1993A&A...279..298H}
{Hummer}, D.G.; {Berrington}, K.A.; {Eissner}, W.; {Pradhan}, A.K.; {Saraph},
  H.E.; {Tully}, J.A.
\newblock {Atomic data from the IRON project. I. Goals and methods.}
\newblock {\em \aap} {\bf 1993}, {\em 279},~298--309.

\bibitem[{Badnell}(1999)]{1999JPhB...32.5583B}
{Badnell}, N.R.
\newblock {A perturbative approach to the coupled outer-region equations for
  the electron-impact excitation of neutral atoms}.
\newblock {\em Journal of Physics B Atomic Molecular Physics} {\bf 1999}, {\em
  32},~5583--5591.
\newblock
  doi:{\changeurlcolor{black}\href{https://doi.org/10.1088/0953-4075/32/23/312}{\detokenize{10.1088/0953-4075/32/23/312}}}.

\bibitem[Badnell(2011)]{badnell:11}
Badnell, N.R.
\newblock A Breit-Pauli distorted wave implementation for AUTOSTRUCTURE.
\newblock {\em Comput. Phys. Commun.} {\bf 2011}, {\em 182},~1528--1535.
\newblock
  doi:{\changeurlcolor{black}\href{https://doi.org/10.1016/j.cpc.2011.03.023}{\detokenize{10.1016/j.cpc.2011.03.023}}}.

\bibitem[{Chidichimo} \em{et~al.}(2003){Chidichimo}, {Badnell}, and
  {Tully}]{2003A&A...401.1177C}
{Chidichimo}, M.C.; {Badnell}, N.R.; {Tully}, J.A.
\newblock {Atomic data from the IRON Project. LII. Electron excitation of
  Ni$^{+24}$}.
\newblock {\em \aap} {\bf 2003}, {\em 401},~1177--1183.
\newblock
  doi:{\changeurlcolor{black}\href{https://doi.org/10.1051/0004-6361:20030131}{\detokenize{10.1051/0004-6361:20030131}}}.

\bibitem[{Chidichimo} \em{et~al.}(2005){Chidichimo}, {Del Zanna}, {Mason},
  {Badnell}, {Tully}, and {Berrington}]{2005A&A...430..331C}
{Chidichimo}, M.C.; {Del Zanna}, G.; {Mason}, H.E.; {Badnell}, N.R.; {Tully},
  J.A.; {Berrington}, K.A.
\newblock {Atomic data from the IRON Project = 2, 3, 4 configurations. LVI.
  Electron excitation of Be-like Fe XXIII for the n = 2,3,4 configurations}.
\newblock {\em \aap} {\bf 2005}, {\em 430},~331--341.
\newblock
  doi:{\changeurlcolor{black}\href{https://doi.org/10.1051/0004-6361:20041358}{\detokenize{10.1051/0004-6361:20041358}}}.

\bibitem[{Del Zanna} and {Mason}(2018)]{delzanna_mason:2018}
{Del Zanna}, G.; {Mason}, H.E.
\newblock {Solar UV and X-ray spectral diagnostics}.
\newblock {\em Living Reviews in Solar Physics} {\bf 2018}, {\em 15},~5.
\newblock
  doi:{\changeurlcolor{black}\href{https://doi.org/10.1007/s41116-018-0015-3}{\detokenize{10.1007/s41116-018-0015-3}}}.

\bibitem[{Badnell} \em{et~al.}(2016){Badnell}, {Del Zanna},
  {Fern{\'a}ndez-Menchero}, {Giunta}, {Liang}, {Mason}, and
  {Storey}]{badnell_etal:2016}
{Badnell}, N.R.; {Del Zanna}, G.; {Fern{\'a}ndez-Menchero}, L.; {Giunta}, A.S.;
  {Liang}, G.Y.; {Mason}, H.E.; {Storey}, P.J.
\newblock {Atomic processes for astrophysical plasmas}.
\newblock {\em Journal of Physics B Atomic Molecular Physics} {\bf 2016}, {\em
  49},~094001.
\newblock
  doi:{\changeurlcolor{black}\href{https://doi.org/10.1088/0953-4075/49/9/094001}{\detokenize{10.1088/0953-4075/49/9/094001}}}.

\bibitem[{Del Zanna} \em{et~al.}(2015){Del Zanna}, {Dere}, {Young}, {Landi},
  and {Mason}]{2015A&A...582A..56D}
{Del Zanna}, G.; {Dere}, K.P.; {Young}, P.R.; {Landi}, E.; {Mason}, H.E.
\newblock {CHIANTI - An atomic database for emission lines. Version 8}.
\newblock {\em \aap} {\bf 2015}, {\em 582},~A56.
\newblock
  doi:{\changeurlcolor{black}\href{https://doi.org/10.1051/0004-6361/201526827}{\detokenize{10.1051/0004-6361/201526827}}}.

\bibitem[{Del Zanna} \em{et~al.}(2021){Del Zanna}, {Dere}, {Young}, and
  {Landi}]{chianti_v10}
{Del Zanna}, G.; {Dere}, K.P.; {Young}, P.R.; {Landi}, E.
\newblock {CHIANTI{\textemdash}An Atomic Database for Emission Lines. XVI.
  Version 10, Further Extensions}.
\newblock {\em \apj} {\bf 2021}, {\em 909},~38.
\newblock
  doi:{\changeurlcolor{black}\href{https://doi.org/10.3847/1538-4357/abd8ce}{\detokenize{10.3847/1538-4357/abd8ce}}}.

\bibitem[{Ljepojevic} and {Burgess}(1990)]{ljepojevic_burgess:1990}
{Ljepojevic}, N.N.; {Burgess}, A.
\newblock {Calculation of the electron velocity distribution function in a
  plasma slab with large temperature and density gradients}.
\newblock {\em Proceedings of the Royal Society of London Series A} {\bf 1990},
  {\em 428},~71--111.
\newblock
  doi:{\changeurlcolor{black}\href{https://doi.org/10.1098/rspa.1990.0026}{\detokenize{10.1098/rspa.1990.0026}}}.

\bibitem[{Dud{\'\i}k} \em{et~al.}(2017){Dud{\'\i}k},
  {Dzif{\v{c}}{\'a}kov{\'a}}, {Meyer-Vernet}, {Del Zanna}, {Young}, {Giunta},
  {Sylwester}, {Sylwester}, {Oka}, and {Mason}]{dudik_etal:2017_review}
{Dud{\'\i}k}, J.; {Dzif{\v{c}}{\'a}kov{\'a}}, E.; {Meyer-Vernet}, N.; {Del
  Zanna}, G.; {Young}, P.R.; {Giunta}, A.; {Sylwester}, B.; {Sylwester}, J.;
  {Oka}, M.; {Mason}, H.E.
\newblock {Nonequilibrium Processes in the Solar Corona, Transition Region,
  Flares, and Solar Wind (Invited Review)}.
\newblock {\em \solphys} {\bf 2017}, {\em 292},~100.
\newblock
  doi:{\changeurlcolor{black}\href{https://doi.org/10.1007/s11207-017-1125-0}{\detokenize{10.1007/s11207-017-1125-0}}}.

\bibitem[{L{\"o}rin{\v{c}}{\'\i}k} \em{et~al.}(2020){L{\"o}rin{\v{c}}{\'\i}k},
  {Dud{\'\i}k}, {Del Zanna}, {Dzif{\v{c}}{\'a}kov{\'a}}, and
  {Mason}]{juraj_etal:2020}
{L{\"o}rin{\v{c}}{\'\i}k}, J.; {Dud{\'\i}k}, J.; {Del Zanna}, G.;
  {Dzif{\v{c}}{\'a}kov{\'a}}, E.; {Mason}, H.E.
\newblock {Plasma Diagnostics from Active Region and Quiet-Sun Spectra Observed
  by Hinode/EIS: Quantifying the Departures from a Maxwellian Distribution}.
\newblock {\em \apj} {\bf 2020}, {\em 893},~34.
\newblock
  doi:{\changeurlcolor{black}\href{https://doi.org/10.3847/1538-4357/ab8010}{\detokenize{10.3847/1538-4357/ab8010}}}.

\bibitem[{Del Zanna} \em{et~al.}(2022){Del Zanna}, {Polito}, {Dud{\'\i}k},
  {Testa}, {Mason}, and
  {Dzif{\v{c}}{\'a}kov{\'a}}]{delzanna_etal:2022_eis_iris}
{Del Zanna}, G.; {Polito}, V.; {Dud{\'\i}k}, J.; {Testa}, P.; {Mason}, H.E.;
  {Dzif{\v{c}}{\'a}kov{\'a}}, E.
\newblock {Diagnostics of Non-Maxwellian Electron Distributions in Solar Active
  Regions from Fe XII Lines Observed by the Hinode Extreme Ultraviolet Imaging
  Spectrometer and Interface Region Imaging Spectrograph}.
\newblock {\em \apj} {\bf 2022}, {\em 930},~61.
\newblock
  doi:{\changeurlcolor{black}\href{https://doi.org/10.3847/1538-4357/ac6174}{\detokenize{10.3847/1538-4357/ac6174}}}.

\bibitem[{Nicholls} \em{et~al.}(2012){Nicholls}, {Dopita}, and
  {Sutherland}]{nicholls_etal:2012}
{Nicholls}, D.C.; {Dopita}, M.A.; {Sutherland}, R.S.
\newblock {Resolving the Electron Temperature Discrepancies in H II Regions and
  Planetary Nebulae: {$\kappa$}-distributed Electrons}.
\newblock {\em \apj} {\bf 2012}, {\em 752},~148.
\newblock
  doi:{\changeurlcolor{black}\href{https://doi.org/10.1088/0004-637X/752/2/148}{\detokenize{10.1088/0004-637X/752/2/148}}}.

\bibitem[{Storey} and {Sochi}(2015)]{storey_sochi:2015b}
{Storey}, P.J.; {Sochi}, T.
\newblock {Effective collision strengths for excitation and de-excitation of
  nebular [O III] optical and infrared lines with {$\kappa$} distributed
  electron energies}.
\newblock {\em \mnras} {\bf 2015}, {\em 449},~2974--2979.
\newblock
  doi:{\changeurlcolor{black}\href{https://doi.org/10.1093/mnras/stv484}{\detokenize{10.1093/mnras/stv484}}}.

\bibitem[{Griffin} \em{et~al.}(1998){Griffin}, {Badnell}, and
  {Pindzola}]{griffin_etal:98}
{Griffin}, D.C.; {Badnell}, N.R.; {Pindzola}, M.S.
\newblock {$R$-matrix electron-impact excitation cross sections in intermediate
  coupling: an MQDT transformation approach}.
\newblock {\em Journal of Physics B Atomic Molecular Physics} {\bf 1998}, {\em
  31},~3713--3727.

\bibitem[{Berrington} \em{et~al.}(1995){Berrington}, {Eissner}, and
  {Norrington}]{berrington_etal:95}
{Berrington}, K.A.; {Eissner}, W.B.; {Norrington}, P.H.
\newblock {RMATRX1: Belfast atomic R-matrix codes}.
\newblock {\em Computer Physics Communications} {\bf 1995}, {\em 92},~290--420.

\bibitem[{Zatsarinny}(2006)]{zatsarinny:2006}
{Zatsarinny}, O.
\newblock {BSR: B-spline atomic R-matrix codes}.
\newblock {\em Computer Physics Communications} {\bf 2006}, {\em
  174},~273--356.
\newblock
  doi:{\changeurlcolor{black}\href{https://doi.org/10.1016/j.cpc.2005.10.006}{\detokenize{10.1016/j.cpc.2005.10.006}}}.

\bibitem[{Norrington} and {Grant}(1981)]{norrington_grant:1981}
{Norrington}, P.H.; {Grant}, I.P.
\newblock {Electron scattering from Ne II using the relativistic R-matrix
  method}.
\newblock {\em Journal of Physics B Atomic Molecular Physics} {\bf 1981}, {\em
  14},~L261--L267.
\newblock
  doi:{\changeurlcolor{black}\href{https://doi.org/10.1088/0022-3700/14/7/006}{\detokenize{10.1088/0022-3700/14/7/006}}}.

\bibitem[{Fern{\'a}ndez-Menchero} \em{et~al.}(2017){Fern{\'a}ndez-Menchero},
  {Zatsarinny}, and {Bartschat}]{fernandez-menchero_etal:2017_n_4}
{Fern{\'a}ndez-Menchero}, L.; {Zatsarinny}, O.; {Bartschat}, K.
\newblock {Electron impact excitation of N$^{3+}$ using the B-spline R-matrix
  method: importance of the target structure description and the size of the
  close-coupling expansion}.
\newblock {\em Journal of Physics B Atomic Molecular Physics} {\bf 2017}, {\em
  50},~065203.
\newblock
  doi:{\changeurlcolor{black}\href{https://doi.org/10.1088/1361-6455/aa5fc4}{\detokenize{10.1088/1361-6455/aa5fc4}}}.

\bibitem[{Del Zanna} \em{et~al.}(2019){Del Zanna}, {Fern{\'a}ndez-Menchero},
  and {Badnell}]{delzanna_etal:2019_n_4}
{Del Zanna}, G.; {Fern{\'a}ndez-Menchero}, L.; {Badnell}, N.R.
\newblock {Uncertainties on atomic data. A case study: N IV}.
\newblock {\em \mnras} {\bf 2019}, {\em 484},~4754--4759.
\newblock
  doi:{\changeurlcolor{black}\href{https://doi.org/10.1093/mnras/stz206}{\detokenize{10.1093/mnras/stz206}}}.

\bibitem[{Burgess} \em{et~al.}(1997){Burgess}, {Chidichimo}, and
  {Tully}]{burgess_etal:97}
{Burgess}, A.; {Chidichimo}, M.C.; {Tully}, J.A.
\newblock {High-energy Born collision strengths for optically forbidden
  transitions }.
\newblock {\em Journal of Physics B Atomic Molecular Physics} {\bf 1997}, {\em
  30},~33--57.

\bibitem[{Chidichimo} \em{et~al.}(2003){Chidichimo}, {Badnell}, and
  {Tully}]{chidichimo_etal:03}
{Chidichimo}, M.C.; {Badnell}, N.R.; {Tully}, J.A.
\newblock {Atomic data from the IRON Project. LII. Electron excitation of
  Ni$^{+24}$}.
\newblock {\em \aap} {\bf 2003}, {\em 401},~1177--1183.

\bibitem[{Burgess} and {Tully}(1992)]{burgess_tully:92}
{Burgess}, A.; {Tully}, J.A.
\newblock {On the Analysis of Collision Strengths and Rate Coefficients}.
\newblock {\em \aap} {\bf 1992}, {\em 254},~436--+.

\bibitem[{Whiteford} \em{et~al.}(2001){Whiteford}, {Badnell}, {Ballance},
  {O'Mullane}, {Summers}, and {Thomas}]{whiteford_etal:01}
{Whiteford}, A.D.; {Badnell}, N.R.; {Ballance}, C.P.; {O'Mullane}, M.G.;
  {Summers}, H.P.; {Thomas}, A.L.
\newblock {A radiation-damped R-matrix approach to the electron-impact
  excitation of helium-like ions for diagnostic application to fusion and
  astrophysical plasmas}.
\newblock {\em Journal of Physics B Atomic Molecular Physics} {\bf 2001}, {\em
  34},~3179--3191.
\newblock
  doi:{\changeurlcolor{black}\href{https://doi.org/10.1088/0953-4075/34/15/320}{\detokenize{10.1088/0953-4075/34/15/320}}}.

\bibitem[{Mao} \em{et~al.}(2022){Mao}, {Del Zanna}, {Gu}, {Zhang}, and
  {Badnell}]{mao_etal_hlike:2024}
{Mao}, J.; {Del Zanna}, G.; {Gu}, L.; {Zhang}, C.Y.; {Badnell}, N.R.
\newblock {R-matrix Electron-impact Excitation Data for the H- and He-like Ions
  with Z = 6-30}.
\newblock {\em \apjs} {\bf 2022}, {\em 263},~35.
\newblock
  doi:{\changeurlcolor{black}\href{https://doi.org/10.3847/1538-4365/ac9c57}{\detokenize{10.3847/1538-4365/ac9c57}}}.

\bibitem[{Gorczyca} and {Badnell}(1996)]{gorczyca_badnell:1996}
{Gorczyca}, T.W.; {Badnell}, N.R.
\newblock {LETTER TO THE EDITOR: Radiation damping in highly charged ions: an
  R-matrix approach}.
\newblock {\em Journal of Physics B Atomic Molecular Physics} {\bf 1996}, {\em
  29},~L283--L290.
\newblock
  doi:{\changeurlcolor{black}\href{https://doi.org/10.1088/0953-4075/29/7/007}{\detokenize{10.1088/0953-4075/29/7/007}}}.

\bibitem[{Griffin} and {Ballance}(2009)]{griffin_ballance:2009}
{Griffin}, D.C.; {Ballance}, C.P.
\newblock {Relativistic radiatively damped R-matrix calculations of the
  electron-impact excitation of helium-like iron and krypton}.
\newblock {\em Journal of Physics B Atomic Molecular Physics} {\bf 2009}, {\em
  42},~235201.
\newblock
  doi:{\changeurlcolor{black}\href{https://doi.org/10.1088/0953-4075/42/23/235201}{\detokenize{10.1088/0953-4075/42/23/235201}}}.

\bibitem[{Fern{\'a}ndez-Menchero} \em{et~al.}(2016){Fern{\'a}ndez-Menchero},
  {Del Zanna}, and {Badnell}]{fernandez-menchero_etal:2016_h_he}
{Fern{\'a}ndez-Menchero}, L.; {Del Zanna}, G.; {Badnell}, N.R.
\newblock {Scaling of collision strengths for highly-excited states of ions of
  the H- and He-like sequences}.
\newblock {\em \aap} {\bf 2016}, {\em 592},~A135.
\newblock
  doi:{\changeurlcolor{black}\href{https://doi.org/10.1051/0004-6361/201628484}{\detokenize{10.1051/0004-6361/201628484}}}.

\bibitem[{Liang} and {Badnell}(2011)]{liang_badnell:2011}
{Liang}, G.Y.; {Badnell}, N.R.
\newblock {R-matrix electron-impact excitation data for the Li-like
  iso-electronic sequence including Auger and radiation damping}.
\newblock {\em \aap} {\bf 2011}, {\em 528},~A69.
\newblock
  doi:{\changeurlcolor{black}\href{https://doi.org/10.1051/0004-6361/201016417}{\detokenize{10.1051/0004-6361/201016417}}}.

\bibitem[{Del Zanna} \em{et~al.}(2015){Del Zanna}, {Dere}, {Young}, {Landi},
  and {Mason}]{delzanna_chianti_v8}
{Del Zanna}, G.; {Dere}, K.P.; {Young}, P.R.; {Landi}, E.; {Mason}, H.E.
\newblock {CHIANTI - An atomic database for emission lines. Version 8}.
\newblock {\em \aap} {\bf 2015}, {\em 582},~A56.
\newblock
  doi:{\changeurlcolor{black}\href{https://doi.org/10.1051/0004-6361/201526827}{\detokenize{10.1051/0004-6361/201526827}}}.

\bibitem[Del~Zanna \em{et~al.}(2008)Del~Zanna, Rozum, and
  Badnell]{delzanna_etal:08_mg_9}
Del~Zanna, G.; Rozum, I.; Badnell, N.
\newblock Electron-impact excitation of Be-like Mg.
\newblock {\em \aap} {\bf 2008}, {\em 487},~1203--1208.
\newblock
  doi:{\changeurlcolor{black}\href{https://doi.org/10.1051/0004-6361:200809998}{\detokenize{10.1051/0004-6361:200809998}}}.

\bibitem[{Chidichimo} \em{et~al.}(2005){Chidichimo}, {Del Zanna}, {Mason}, and
  {et al.}]{chidichimo_etal:05}
{Chidichimo}, M.C.; {Del Zanna}, G.; {Mason}, H.E.; {et al.}.
\newblock {Atomic Data from the IRON Project LVI. Electron excitation of
  Be-like Fe~{XXIII}}.
\newblock {\em \aap} {\bf 2005}, {\em 430},~331.

\bibitem[{Fern{\'a}ndez-Menchero} \em{et~al.}(2014){Fern{\'a}ndez-Menchero},
  {Del Zanna}, and {Badnell}]{fernandez-menchero_etal:2015_be-like}
{Fern{\'a}ndez-Menchero}, L.; {Del Zanna}, G.; {Badnell}, N.R.
\newblock {R-matrix electron-impact excitation data for the Be-like
  iso-electronic sequence}.
\newblock {\em \aap} {\bf 2014}, {\em 566},~A104.
\newblock
  doi:{\changeurlcolor{black}\href{https://doi.org/10.1051/0004-6361/201423864}{\detokenize{10.1051/0004-6361/201423864}}}.

\bibitem[Liang \em{et~al.}(2012)Liang, Badnell, and Zhao]{liang_etal:2012}
Liang, G.Y.; Badnell, N.R.; Zhao, G.
\newblock R-matrix electron-impact excitation data for the B-like
  iso-electronic sequence.
\newblock {\em \aap} {\bf 2012}, {\em 547},~A87.
\newblock
  doi:{\changeurlcolor{black}\href{https://doi.org/10.1051/0004-6361/201220277}{\detokenize{10.1051/0004-6361/201220277}}}.

\bibitem[{Badnell} \em{et~al.}(2001){Badnell}, {Griffin}, and
  {Mitnik}]{badnell_etal:01}
{Badnell}, N.R.; {Griffin}, D.C.; {Mitnik}, D.M.
\newblock {Electron-impact excitation of Fe$^{21+}$, including n = 4 levels }.
\newblock {\em Journal of Physics B Atomic Molecular Physics} {\bf 2001}, {\em
  34},~5071--5085.

\bibitem[{Fern{\'a}ndez-Menchero} \em{et~al.}(2016){Fern{\'a}ndez-Menchero},
  {Giunta}, {Del Zanna}, and {Badnell}]{menchero_etal:2016_fe21}
{Fern{\'a}ndez-Menchero}, L.; {Giunta}, A.S.; {Del Zanna}, G.; {Badnell}, N.R.
\newblock {Importance of the completeness of the configuration interaction and
  close coupling expansions in R-matrix calculations for highly charged ions:
  electron-impact excitation of Fe$^{20+}$}.
\newblock {\em Journal of Physics B Atomic Molecular Physics} {\bf 2016}, {\em
  49},~085203.
\newblock
  doi:{\changeurlcolor{black}\href{https://doi.org/10.1088/0953-4075/49/8/085203}{\detokenize{10.1088/0953-4075/49/8/085203}}}.

\bibitem[{Mao} \em{et~al.}(2020){Mao}, {Badnell}, and {Del
  Zanna}]{mao_etal:2020}
{Mao}, J.; {Badnell}, N.R.; {Del Zanna}, G.
\newblock {R-matrix electron-impact excitation data for the C-like
  iso-electronic sequence}.
\newblock {\em \aap} {\bf 2020}, {\em 634},~A7.
\newblock
  doi:{\changeurlcolor{black}\href{https://doi.org/10.1051/0004-6361/201936931}{\detokenize{10.1051/0004-6361/201936931}}}.

\bibitem[{Witthoeft} \em{et~al.}(2007){Witthoeft}, {Whiteford}, and
  {Badnell}]{witthoeft_etal:2007}
{Witthoeft}, M.C.; {Whiteford}, A.D.; {Badnell}, N.R.
\newblock {R-matrix electron-impact excitation calculations along the F-like
  iso-electronic sequence}.
\newblock {\em Journal of Physics B Atomic Molecular Physics} {\bf 2007}, {\em
  40},~2969--2993.
\newblock
  doi:{\changeurlcolor{black}\href{https://doi.org/10.1088/0953-4075/40/15/001}{\detokenize{10.1088/0953-4075/40/15/001}}}.

\bibitem[{Mao} \em{et~al.}(2020){Mao}, {Badnell}, and {Del
  Zanna}]{mao_etal:2020_n-like}
{Mao}, J.; {Badnell}, N.R.; {Del Zanna}, G.
\newblock {R-matrix electron-impact excitation data for the N-like
  iso-electronic sequence}.
\newblock {\em \aap} {\bf 2020}, {\em 643},~A95.
\newblock
  doi:{\changeurlcolor{black}\href{https://doi.org/10.1051/0004-6361/202039195}{\detokenize{10.1051/0004-6361/202039195}}}.

\bibitem[{Butler} and {Badnell}(2008)]{butler_badnell:2008}
{Butler}, K.; {Badnell}, N.R.
\newblock {Atomic data from the IRON project. LXVI. Electron impact excitation
  of Fe$^{18+}$}.
\newblock {\em \aap} {\bf 2008}, {\em 489},~1369--1376.
\newblock
  doi:{\changeurlcolor{black}\href{https://doi.org/10.1051/0004-6361:200810197}{\detokenize{10.1051/0004-6361:200810197}}}.

\bibitem[{Mao} \em{et~al.}(2021){Mao}, {Badnell}, and {Del
  Zanna}]{mao_etal:2021_o-like}
{Mao}, J.; {Badnell}, N.R.; {Del Zanna}, G.
\newblock {R-matrix electron-impact excitation data for the O-like
  iso-electronic sequence}.
\newblock {\em \aap} {\bf 2021}, {\em 653},~A81.
\newblock
  doi:{\changeurlcolor{black}\href{https://doi.org/10.1051/0004-6361/202141464}{\detokenize{10.1051/0004-6361/202141464}}}.

\bibitem[Liang and Badnell(2010)]{liang_badnell:10_ne-like}
Liang, G.Y.; Badnell, N.R.
\newblock R-matrix electron-impact excitation data for the Ne-like
  iso-electronic sequence.
\newblock {\em \aap} {\bf 2010}, {\em 518},~A64.
\newblock
  doi:{\changeurlcolor{black}\href{https://doi.org/10.1051/0004-6361/201014170}{\detokenize{10.1051/0004-6361/201014170}}}.

\bibitem[{Del Zanna}(2011)]{delzanna:2011_fe_17}
{Del Zanna}, G.
\newblock {Benchmarking atomic data for astrophysics: Fe XVII X-ray lines}.
\newblock {\em \aap} {\bf 2011}, {\em 536},~A59.
\newblock
  doi:{\changeurlcolor{black}\href{https://doi.org/10.1051/0004-6361/201117287}{\detokenize{10.1051/0004-6361/201117287}}}.

\bibitem[Loch \em{et~al.}(2006)Loch, Pindzola, Ballance, and
  Griffin]{loch_etal:06}
Loch, S.D.; Pindzola, M.S.; Ballance, C.P.; Griffin, D.C.
\newblock The effects of radiative cascades on the x-ray diagnostic lines of
  Fe$^{16+}$.
\newblock {\em Journal of Physics B Atomic Molecular Physics} {\bf 2006}, {\em
  39},~85--104.
\newblock
  doi:{\changeurlcolor{black}\href{https://doi.org/10.1088/0953-4075/39/1/009}{\detokenize{10.1088/0953-4075/39/1/009}}}.

\bibitem[{Liang} \em{et~al.}(2009{\natexlab{a}}){Liang}, {Whiteford}, and
  {Badnell}]{liang_etal:09_na-like}
{Liang}, G.Y.; {Whiteford}, A.D.; {Badnell}, N.R.
\newblock {R-matrix electron-impact excitation data for the Na-like
  iso-electronic sequence}.
\newblock {\em \aap} {\bf 2009}, {\em 500},~1263--1269.
\newblock
  doi:{\changeurlcolor{black}\href{https://doi.org/10.1051/0004-6361/200911866}{\detokenize{10.1051/0004-6361/200911866}}}.

\bibitem[{Liang} \em{et~al.}(2009{\natexlab{b}}){Liang}, {Whiteford}, and
  {Badnell}]{liang_etal:09_na-like_inner}
{Liang}, G.Y.; {Whiteford}, A.D.; {Badnell}, N.R.
\newblock {R-matrix inner-shell electron-impact excitation of the Na-like
  iso-electronic sequence}.
\newblock {\em Journal of Physics B Atomic Molecular Physics} {\bf 2009}, {\em
  42},~225002.
\newblock
  doi:{\changeurlcolor{black}\href{https://doi.org/10.1088/0953-4075/42/22/225002}{\detokenize{10.1088/0953-4075/42/22/225002}}}.

\bibitem[Fern{\'a}ndez-Menchero \em{et~al.}(2014)Fern{\'a}ndez-Menchero,
  Del~Zanna, and Badnell]{fernandez-menchero_etal:2014_mg-like}
Fern{\'a}ndez-Menchero, L.; Del~Zanna, G.; Badnell, N.R.
\newblock R-matrix electron-impact excitation data for the Mg-like
  iso-electronic sequence.
\newblock {\em \aap} {\bf 2014}, {\em 572},~A115.
\newblock
  doi:{\changeurlcolor{black}\href{https://doi.org/10.1051/0004-6361/201424849}{\detokenize{10.1051/0004-6361/201424849}}}.

\end{thebibliography}

%%%%%%%%%%%%%%%%%%%%%%%%%%%%%%%%%%%%%%%%%%
\end{document}